\documentclass[11pt,a4paper]{article}

\usepackage[suppress]{color-edits}
\addauthor{AMM}{teal}
\addauthor{SO}{magenta}
\addauthor{SG}{blue}
\addauthor{YJ}{purple}
\addauthor{PA}{orange}
\addauthor{JDR}{olive}

\usepackage[hyperref]{acl2021}
\usepackage{times}
\usepackage{graphicx}
\usepackage{latexsym}

\usepackage{booktabs}

\usepackage{microtype}

\aclfinalcopy 
\def\aclpaperid{8} 

\title{Reusable Templates and Guides For Documenting Datasets and Models \\for Natural Language Processing and Generation\\
\vspace{0.5em}\large{A Case Study of the HuggingFace and GEM Data and Model Cards}}

\author{Angelina~McMillan-Major\\University of Washington\\Hugging Face\\\texttt{angie@huggingface.co} 
\And Salomey~Osei\\KNUST\\Masakhane\\\texttt{sosei@aimsammi.org} 
\And Juan Diego~Rodriguez\\UT Austin\\\texttt{juand-r@utexas.edu} \\
\AND Pawan~Sasanka~Ammanamanchi\\IIIT Hyderabad\\\texttt{pass2pawan@gmail.com} 
\And Sebastian Gehrmann\\Google Research\\\texttt{gehrmann@google.com} 
\And Yacine Jernite\\Hugging Face\\\texttt{yacine@huggingface.co}}

\date{}

\begin{document}
\maketitle
\begin{abstract}
Developing documentation guidelines and easy-to-use templates for datasets and models is a challenging task, especially given the variety of backgrounds, skills, and incentives of the people involved in the building of natural language processing (NLP) tools.
Nevertheless, the adoption of standard documentation practices across the field of NLP promotes more accessible and detailed descriptions of NLP datasets and models, while supporting researchers and developers in reflecting on their work.
To help with the standardization of documentation, we present two case studies of efforts that aim to develop reusable documentation templates -- the HuggingFace data card, a general purpose card for datasets in NLP, and the GEM benchmark data and model cards with a focus on natural language generation.
We describe our process for developing these templates, including the identification of relevant stakeholder groups, the definition of a set of guiding principles, the use of existing templates as our foundation, and iterative revisions based on feedback. 
\end{abstract}

\section{Introduction}

Dataset and model documentation is a necessary step in identifying potential issues with machine learning (ML) systems and addressing their broader impacts~\citep[][among others]{gebru2018datasheets,gebru2020datasheets, bender-friedman-2018-data}.
In their overview of data collection and use in ML, \citet{paullada2020data} identify issues that have frequently arisen such as considerations for how subjects are represented in datasets, spurious cues that may be exploited by ML model, and concerns about the content in datasets collected through crawling methodologies. They advocate for careful documentation of datasets and their collection processes in order to surface these problems. However, best practices for documentation have seen no widespread adoption even for the most popular datasets and models. Indeed, writing such detailed documentation requires additional effort from researchers who may lack the required resources or familiarity with the process. 
Providing dataset and model creators with guidelines and several examples in a single place to inspire and inform prospective writers could thus drive widespread adoption of documentation.

Research efforts that involve a large number of models or datasets are particularly well positioned to develop and maintain specific guidelines and best practices by making documentation a required component for submitting contributions. 
By bringing together the domain expertise of participants and the experience of researchers who have a greater familiarity with documenting data and models, these efforts provide an opportunity to develop and refine templates that balance generality and informativeness.
In addition, by requiring appropriate documentation for any involved model or dataset, these efforts can set a precedent that informs future endeavours.
We encourage organizations to consider their role in the successful uptake of documentation practices, such as providing their members with adequate resources to understand the goals and motivations of documentation and measured steps towards integrating documentation into current research norms.

In this paper, we present two case studies of creating documentation templates and guides in natural language processing (NLP): the Hugging Face (HF) dataset hub\footnote{\url{https://hf.co/datasets/card-guide}} and the benchmark for Generation and its Evaluation and Metrics (GEM).\footnote{\url{https://gem-benchmark.com/data\_cards/}} 
We use the term \textit{data card} to refer to documentation for datasets in both cases and the term \textit{model card} to refer to documentation for models in the GEM workshop, following \citet{mitchell2019modelcards}.
Focusing on these settings allows us to ground what constitutes `good' documentation in these contexts, namely technical user-oriented information, scientific reproducibility, and social contextualization of data and data-driven systems.

\section{Related Work} \label{sec:relatedwork}

In the U.S., research involving direct interventions on human subjects is subject to the Federal Policy for the Protection of Human Subjects (or Common Rule).\footnote{\url{https://www.nidcr.nih.gov/research/human-subjects-research}}
This policy tasks institutional review boards (IRBs) with certifying that such research follows established ethical standards and regulations. While this review is sometimes decried as cumbersome~\citep{grady2015institutional}, the process ensures that researchers both reflect and communicate ahead of time how the data will be collected and used, why it is necessary to answer the question at hand, and how protected information will be handled to prevent harm to the human subjects.
Whereas much of the data that supports current methods in ML and especially NLP is created by, gives information about, and is used to train models that will likely affect these same human subjects, this relationship does not constitute a direct intervention as defined in the Common Rule. However, \citeauthor{Metcalf2016WhereAH} argue that the definition is too narrow when one considers the similarity in potential harms and advise that data-driven methods should be subject to a similar from of ethical review, which includes clear communication about the goals and mechanisms for collecting and safeguarding the data.

Despite existing literature on database documentation in HCI and related fields \citep{cheney2009provenance, bhardwaj2014datahub}, documentation in ML has only recently gained traction. In a survey of ML projects across India, East and West African countries, and the USA, \citet{sambasivan2021datacascades} analyze compounding events causing negative, downstream effects from data issues, resulting in technical debt\footnote{\citet{10.1145/157709.157715} employs this term to describe the accumulation of flaws in technical systems over time, using an analogy to financial debt.} over time, and identify insufficient documentation to be a trigger of these events in 20\% of the 53 cases. 
They report impacts such as time and effort lost in using incorrect data, barriers to completing models, and data that had to be abandoned due to it no longer being usable.

In response to issues like the ones found by \citeauthor{sambasivan2021datacascades}, many research groups have proposed documentation schemata for different parts of the ML pipeline. \citet{arnold2019factsheets} introduce FactSheets to document the use, performance and security aspects of an AI product. \citet{mitchell2019modelcards} put forward Model Cards focusing on documenting the evaluation and use of a specific model, while \citet{gebru2018datasheets}, \citet{holland2018dataset}, and \citet{pushkarna2021datacardsplaybook} propose Datasheets for Datasets, Dataset Nutrition Labels, and the Data Cards Playbook, respectively, as documentation schemata and processes for documenting the data used in ML and AI systems. \citet{hutchinson2021datasets} frame datasets as technical infrastructure and propose documentation for several stages of the development process, including the design, creation, and maintenance of the dataset.

To address the distinct challenges of working with language data, such as those summarized by \citet{10.1145/3442188.3445922b}, other researchers have proposed specialized documentation for work in NLP. The first such example is by \citet{bender-friedman-2018-data} who propose a version of Data Statements for documenting aspects of the data from a linguistic perspective. In addition to documenting the data that a model sees during training, there is also the need to document experiments, especially those involving humans. Thus, \citet{shimorina2021heds} develop a Human Evaluation Datasheet, with the goal of describing human evaluation experiments rather than naturally occurring language data. 
Starting from the templates by \citet{bender-friedman-2018-data} and \citet{mitchell2019modelcards}, we iteratively extend and adapt our templates for the HF and GEM contexts.
\section{Methods} \label{sec:methods}
To guide the development of our templates, we draw from the methods of value sensitive design \citep[eg.,][]{friedman2019value} by identifying stakeholders and assessing their values, which \citeauthor{friedman2019value} define as ``what is important to people in their lives, with a focus on ethics and morality" (pg. 24). The values of the various stakeholders, including developers themselves, may influence the development of a technology in a number of ways \citep{friedmankahn2003humanvalues}. In this section, we conduct a stakeholder analysis and describe the principles we follow in the development of our templates. In Section~\ref{sec:impact-and-positionality}, we explore the potential social impact of the templates and our positionality in the design process.

\subsection{Documentation Stakeholders} \label{sec:methods-stakeholders}

This work presents a documentation strategy adopted by two {\textbf{organizations}} (the HF dataset hub and the GEM benchmark) for two categories of {\textbf{resources}} (language datasets and NLG models). We identify three groups of direct stakeholders as well as indirect stakeholders whose needs we consider in designing this documentation strategy.

\paragraph{The organizations.} The managing organizations play a central role in gathering, presenting, and enabling the use of the resources. The organizations are responsible for \textit{establishing documentation standards} that need to be met for any resource.

\paragraph{The resource creators.} When submitting their resources to an organization, dataset curators and model developers are required to \textit{write documentation} to meet the organization's stated standards.

\paragraph{The resource users.} The resources distributed by the organizations may further be utilized by downstream users. These users \textit{read the provided documentation} to determine whether the resources may or may not be appropriate for their needs.

\paragraph{The indirect stakeholders.} Indirect stakeholders include any person impacted by the dissemination of a resource, such as dataset publication, model deployment, and deployment of a model trained on a dataset. While indirect stakeholders might not have direct control over the way the resources are used, they may \textit{refer to the documentation} to analyze these impacts.

\subsection{General Approach} \label{sec:methods-approach}

As discussed by \citet{waseem2021disembodied}, datasets are the result of subjective choices made by the dataset creators. We therefore approach documentation as a way for authors to communicate this subjectivity by explicitly stating the decisions that led to the resource creation and the contexts in which those decisions were made. In addition to providing developers with the opportunity to reflect on their choices in creating a resource, the documentation gives users insight as to how and why the resource was developed, which may help the user assess how appropriate the resource is for their use case, and may even surface previously unconsidered issues.
In addition to the documentation formats surveyed in Section \ref{sec:relatedwork}, the practice of reflecting on the impact of one's work is being standardized by academic conferences such as NeurIPS's broader impacts statement\footnote{https://neurips.cc/Conferences/2020/CallForPapers} and NAACL's ethics review.\footnote{https://2021.naacl.org/ethics/faq/} We aim to encourage this practice in our local contexts through our free text templates that emphasize reflection on the topics we see as important in understanding the development and potential uses of datasets and models in NLP.

While guidance around these processes is still being standardized, we recognize that there are risks that may result from the proliferation of documentation formats and the suggestion of documentation as a way to mitigate the harms caused by ML systems. 
For example, documentation that is not updated to reflect changes to the documented resource may result in harms due to decisions made on inaccurate information.
In other cases, the standardization of documentation may add to the difficulties experienced by inexperienced or underfunded researchers in publishing their work.
Finally, authors may try to justify work that causes known harms by documenting the potential for harm without attempting to address the harms themselves.
Organizations that institute documentation standards need to consider these risks when integrating documentation into their local contexts and be attentive to the varied impacts to researchers, developers, and community members.

\subsection{Data Cards Principles} \label{sec:methods-datacards}

Language conveys information about not only the individual producing the language, but also about the social groups that individual is a member of and social context that the individual is producing the language in \citep{eckertrickford2001style}. 
For example, an accent may indicate the geographical region that a person grew up in and that person's use of a local phrase may indicate that they believe the person that they're talking to is also from that region.
As such, the values of our indirect stakeholders, the people whose sociolinguistic information are embodied in the resources, are of high priority in designing the data card templates. 
In order for documentation to be accessible to indirect stakeholders, who may not be familiar with ML terminology or academic writing, the information needs to be clearly presented and easy to find within the document. 
This consideration led to several revisions in our template content and presentation.

\subsection{Model Cards Principles} \label{sec:methods-modelcards}

The recent push by academic conferences for social impact statements in publications provides a clear way for resource creators to consider the impact to their own direct and indirect stakeholders. 
Furthermore, academic ML conferences such as NeurIPS have instituted reproducibility checklists for paper submissions \cite{pineau2020improving}. Building off this checklist, \citet{dodge-etal-2019-show} argue for greater reproducibility in ML publications by proposing their own reproducibility checklist as well as a metric for reporting performance on the validation set as a function of the model training time. \citet{mitchell2019modelcards} focus on evaluation in their own schema for models, arguing for disaggregated results to be reported over different population subgroups in data used to evaluate the model. These three considerations - social impact, reproducibility, and evaluation - form the main aspects of our template.

\section{Case Study I: HuggingFace Data Cards} \label{sec:huggingface}

As a first case study of data card development, we present the template developed for the HuggingFace open source NLP libraries.
The full template is available in Table~\ref{tab:data-card-comparison}.
The completed data card for the ELI5 dataset \citep{fan-etal-2019-eli5} is available in Appendix~\ref{sec:appendix-eli5}.

\paragraph{HF Libraries} \YJedit{The HuggingFace Transformers library~\citep{hf_transformers} has evolved as a centralized platform providing a common API for easily loading the weights of nearly 10,000 (at the time of writing) transformer-based models with different frameworks (PyTorch, TensorFlow, JAX) and original code bases. The Datasets library\footnote{\url{https://hf.co/datasets}} takes a similar approach to providing a hub that allows users to easily discover and reuse the datasets that were used to trained those models. To that end, the library focuses on the following three features:
\begin{itemize}
    \item A unified API for downloading and iterating through a wide variety of datasets
    \item A backend supported by memory-mapped Arrow arrays\footnote{\url{https://arrow.apache.org/}} to enable use of even large datasets in resource-constrained settings
    \item A documentation structure that gives users a clear overview of the available datasets and the information required to use them
\end{itemize}
To meet the latter need, we designed a data card template that provides a unified way to present this information for all of the proposed NLP datasets.}

\paragraph{HF data cards stakeholders} The direct stakeholders of the HF card templates include: the organization, whose goals for the library and its documentation standards are stated above; the team and HF community members, who add new datasets to the library and are encouraged to fill out as much of the cards as they can; and the library users, who may examine or train models on the provided datasets. We note that, in our case, the people who add the datasets to the library may not be the original dataset curators, and so may not have direct access to all the required information. 

\paragraph{Initial version}
The first version of the data card consisted of 8 sections. The first three aimed to answer the question \textit{``What is this dataset used for?''} and asked the writers to fill out information about the tasks supported, the original purpose for creating the dataset, and the languages represented within. The template then asked about the \textit{people involved in making the dataset}, including the dataset creators, the language creators, and the annotators, if relevant. This was followed by a section titled ``Data Characteristics," which covered all of the \textit{data selection and processing steps}. In particular, considering that users might want to use several datasets developed to address similar tasks together to train a model, we wanted to surface any domain shifts or differences in text normalization in either of these last two sections. 

The template then continued to a \textit{dataset structure} section covering information about the default train/test split if provided, size of the dataset, description of the features, examples of data points, and suggested metrics to use. Broadly, the purpose of this section was to give a user any technical information they might need to train a model, such as how the dataset size might influence what size of model or regularization technique to use. Seeing the features and an explicit example of a data instance can also clarify the input and output features and texts. 

The second to last section covered \textit{known limitations} of the dataset, and prompted the writer to consider specifically social biases in a first sub-section and any other limitations, such as common surface correlations a model might take advantage of, in another. The reasoning behind this choice is that a user might do as much harm by deploying a model exhibiting harmful biases as by deploying a model that had a high score for the chosen metric but did not actually perform the described high-level task. Finally, for the benefit of users who might want to share derivatives of the dataset or use them for commercial purposes, the last section contained the \textit{licensing information} for the dataset.
Using this schema, we wrote an initial draft card for the SNLI dataset \cite{snli:emnlp2015} and shared it with the HF team and the SNLI curators for comments.

\paragraph{Revised version}
We then expanded and reordered the template based on the initial comments. The first section became the \textit{dataset description} consisting of: a list of links to relevant information about the dataset available elsewhere including the dataset paper, leaderboards, and the contact information of at least one person in case of further questions; a free text summary; a description of supported tasks, suggested metrics (moved here for this updated version), and leaderboards; and the languages represented. The \textit{dataset structure} was moved just after the languages, with examples of data instances and information about the fields and splits.

The largest change was in the people and dataset characteristics sections. We restructured these into a single \textit{dataset creation} section which now starts with a curation rationale to properly contextualize all of the choices described in the rest of the card. The template then requests information about the people involved in producing the source language and annotations and the normalization and processing steps for that data. A section was then added to specifically describe the status of Personal Identifying (PI) data in the dataset in order to both help protect the data subjects' privacy and to help the dataset users comply with existing regulations. Dataset creators were renamed dataset curators to emphasize the difference between the people making the curation choices and the people producing the source language data, and their description was moved to the very end of the data card.
Finally we expanded the limitations section to a broader section on \textit{considerations for using the data}, adding a prompt for prospective social impacts of using the dataset, both positive and negative.

\paragraph{Supporting documentation writers} We made these changes to improve card readers' ability to navigate the document and find necessary information about the dataset and to assist card authors when writing their cards by clarifying the desired information for each section. To further aid authors, we developed a guide formatted with desired content and instructions for each section.\footnote{\url{https://hf.co/datasets/card-guide}} 
We intend for the template and guide to support authors of datasets both with and without existing documentation. Authors of datasets without publications can use the card as a starting point for building documentation and visibility for the dataset in the HF library, but also as an overview of what information should be included in a publication. For datasets with publications, the HF card provides authors with a more widely accessible format for documenting their dataset that does not have the length limitations of paper submissions and can be revised as needed to reflect any updates to the dataset.

Publications also have the property of being static and cannot be updated to reflect changes in the dataset.
To address this, we designed the HF data cards to function as living documents. First, hosting them on GitHub allows community members at large to easily add new information or modify existing sections to reflect new findings. We see this as particularly important for the section on using the dataset as new considerations are reported as a result of novel use cases and research. 
We also made the decision to publish the template with the sections pre-populated with placeholder text (specifically, ``More Information Needed'') in order to encourage authors and community members to fill in the section when the information is available.
The ability to update information helps to address the harms caused by out-of-date documentation. 
By integrating the data cards into the HF library, we are able to see a more complete characterization of the available datasets that is similarly up to date. 
This allows us to point out where fewer datasets are available for tasks and languages and make progress towards a more diverse library. 

\section{Case Study II: GEM Data and Model Cards} \label{sec:gem}

Our second case study of data card development is the template we developed for the datasets and model submissions of the Generation, its Evaluation, and Metrics (GEM) workshop.
We present the full template in Table~\ref{tab:data-card-comparison}.
The completed data card for the ASSET dataset \citep{alva-manchego-etal-2020-asset} is available in Appendix~\ref{sec:appendix-asset}.

\subsection{GEM Benchmark}

The benchmark for GEM aims to standardize how research in natural language generation (NLG) is conducted with a particular focus on in-depth evaluations~\citep{gem-paper}. 
To this end, newly developed NLG models should be documented and evaluated on a set of established tasks over a range of reproducible and robust metrics. This goal can only be achieved if the infrastructure provided by the benchmark supports the creation of such documentation.

Since a benchmark may comprise multiple datasets and provide a centralized way to interact with them, we can focus on two groups of stakeholders following the descriptions in Section~\ref{sec:methods-stakeholders}.
\paragraph{Benchmark curators.} The curators need to \textit{ensure that all datasets are documented} according to the requirements.
\paragraph{Benchmark participants.} The participants need to \textit{write model documentation} according to the requirements, but may be novice ML practitioners or inexperienced in writing documentation.

\subsection{Data Cards} \label{sec:gem-datacards}

{\renewcommand{\arraystretch}{1.18}
\begin{table*}[!tbh]
\centering
\begin{tabular}{@{}ll@{}}
\toprule
\textbf{HuggingFace data card} & \textbf{GEM data card} \\ \midrule
Dataset Description & Dataset and Task Description \\
\textbullet\ Dataset Summary & \textbullet\ Dataset and Task Summary \\
\textbullet\ Supported Tasks and Leaderboards & \textit{See below} \\
-- & \textbullet\ Why is this dataset part of GEM? \\
\textbullet\ Languages & \textbullet\ Languages \\ \midrule
-- & Meta Information \\
\textit{See below}  & \textbullet\ Dataset Curators \\
\textit{See below}  & \textbullet\ Licensing Information \\
\textit{See below}  & \textbullet\ Citation Information \\
\textit{See above}  & \textbullet\ Leaderboard \\ \midrule
Dataset Structure & Dataset Structure \\
\textbullet\ Data Instances & \textbullet\ Data Instances \\
\textbullet\ Data Fields & \textbullet\ Data Fields \\
\textbullet\ Data Splits & \textbullet\ Data Statistics \\ \midrule
Dataset Creation & Dataset Creation \\
\textbullet\ Curation Rationale & \textbullet\ Curation Rationale \\
-- & \textbullet\ Communicative Goal \\
\textbullet\ Source Data & \textbullet\ Source Data \\
\textbullet\textbullet\ Initial Data Collection and Normalization & \textbullet\textbullet\ Initial Data Collection and Normalization \\
\textbullet\textbullet\ Who are the source language producers? & \textbullet\textbullet\ Who are the source language producers? \\
\textbullet\ Annotations & \textbullet\ Annotations \\
\textbullet\textbullet\ Annotation process & \textbullet\textbullet\ Annotation process \\
\textbullet\textbullet\ Who are the annotators? & \textbullet\textbullet\ Who are the annotators? \\
\textbullet\ Personal and Sensitive Information & \textbullet\ Personal and Sensitive Information \\ \midrule
-- & Changes to the Original Dataset for GEM \\ \midrule
Considerations for Using the Data & Considerations for Using the Data \\
\textbullet\ Social Impact of the Dataset & \textbullet\ Social Impact of the Dataset \\
-- & \textbullet\ Impact on Underserved Communities \\
\textbullet\ Discussion of Biases & \textbullet\ Discussion of Biases \\
\textbullet\ Other Known Limitations & \textbullet\ Other Known Limitations \\ \midrule
-- & Getting started with in-depth research on the task \\ \midrule
Additional Information & -- \\
\textbullet\ Dataset Curators & \textit{See above}  \\
\textbullet\ Licensing Information & \textit{See above}  \\
\textbullet\ Citation Information & \textit{See above}  \\
\textbullet\ Contributions & -- \\
-- & Credits for Data Cards and this Template \\ \bottomrule
\end{tabular}
\caption{Side by side comparison of the HF and GEM data card templates. Each section is denoted by horizontal lines, subsections are denoted with \textbullet, subsubsections with \textbullet\textbullet.}
\label{tab:data-card-comparison}
\end{table*}
}

None of the datasets included in GEM had existing data cards. To address this issue, we develop a data card template and use it to document all the datasets involved in the benchmark. Moreover, to be able to quickly add new datasets and to help the broader NLG community construct their own data cards, we release the template and associated guide. The data card closely follows the HF data card template introduced in Section~\ref{sec:huggingface}, with changes to target NLG-specific issues. We made these changes to address feedback after testing an initial version of the data card on the CommonGen~\citep{lin2019commongen} and ASSET~\citep{alva-manchego-etal-2020-asset} datasets. An overview of the differences between the two templates is presented in Table~\ref{tab:data-card-comparison}.

The major difference between the general HF template and the NLG-specific template is that NLG datasets may contain natural language both in the input and the output. Inputs and outputs may have different sources and thus require documentation for both. In addition, the input-output pairs may be constructed in ways that are challenging to describe in the HF template. For example, output text may be crawled and undergo revisions while the input text remains the same. This difference did not lead to different sections in the data card itself, but it did lead to changes in the guidelines on how to write them.

Moreover, NLG tasks have an underlying communicative goal which differentiates them from classification and other structured tasks. It is imperative to surface the communicative goal, since it heavily influences how generated text for a particular task should be evaluated. Another category of changes concerns the context of GEM compared to general purpose data cards. For example, since GEM itself is a benchmark, information about leaderboards does not have to be prominently featured, whereas it should give credit to the original data creators early on.

We also added three GEM-specific sections: (1) Why is this dataset part of GEM, (2) Changes to the original dataset for GEM, and (3) Getting started with in-depth research on the task. The first aims to tie the collections of data cards together by situating a dataset and task within the larger goal of the benchmark. The second section is of crucial importance for any data card for a benchmark, since the benchmark may change the purpose of a dataset and the organizers could modify the underlying data by cleaning it, adding more data, or releasing a reformatted version.
The final question encourages participants to engage with the data in order to develop a deeper understanding of the task formulation.
Therefore, to help participants gain insights into the data, we included a section with helpful pointers to relevant papers and tutorials.

Finally, GEM is designed to be a multilingual benchmark. Since we expect to include languages with fewer resources than may be found for languages like English, we aim to consider the communities that speak those languages and the impacts that technology built with these datasets could have on them. 
For example, a dataset for a language with few other resources may only capture the language of a few speakers in a certain context, like university students in the context of an experiment.
Models trained on this dataset would not have access to the variation in the language that comes from speakers of other ages and in other contexts, but because there may not be other available tools, that model may become widely used and misrepresented as a general model of the language.
We thus added a specific section to address potential concerns involving communities that speak those language varieties, such as the implications for model generality and privacy when speakers from small communities are easily identifiable.

\subsection{Model Cards} \label{sec:gem-modelcards}

\begin{table}[!tb]
\centering
\begin{tabular}{@{}l@{}}
\toprule
\textbf{GEM Model Card} \\ \midrule
Social Impact \\
\textbullet\ Additional Data \\
\textbullet\ Training Process \\
\textbullet\ Real-World Use \\
\textbullet\ Measuring Impact \\ \midrule
Reproducibility \\ 
\textbullet\ Model Description \\
\textbullet\ Model Details \\
\textbullet\ Model Hyperparameters \\
\textbullet\ Hyperparameter Specification \\
\textbullet\ Number of Training and Evaluation Runs \\
\textbullet\ Dataset Details \\
\textbullet\ Dependencies and External Libraries \\
\textbullet\ Link to Downloadable Source Code \\
\textbullet\ Computing Infrastructure Used \\ \midrule
Evaluation details \\ \bottomrule
\end{tabular}
\caption{An overview of the GEM model card template.}
\label{tab:gem-model-card}
\end{table}

Following the guiding principles outlined in Section~\ref{sec:methods-modelcards}, the model cards have three sections: social impact, reproducibility, and evaluation. A detailed overview is shown in Table~\ref{tab:gem-model-card}.

\paragraph{Social impact}
In the first section, we invite submission authors to consider the impacts their models may have on users if they were deployed. We recognize that without further guidance, the open-ended nature of this request may make it prohibitively difficult to address. Indeed, trying to foresee all the ways in which data or modeling choices may affect all direct and indirect stakeholders is overwhelming, if not impossible. Instead, we narrow the scope to help users practice reflecting on causal relationships between design and deployment effects. 
We do this by providing guiding examples of models and their potential impacts. In one scenario, we consider a summarization model trained on a English Wikipedia, which is known to have various dimensions of gender bias \citep{wagner2015s}. We present two possible impacts on the output summaries based on this gender bias and suggest tests to measure the effect.
We then encourage model creators to follow a similar line of thought. We ask about additional data used (and to a link to documentation if it exists), about the training process, and about a possible real-world use. We then request that the documentation author choose one aspect of one of the steps outlined, contemplate a way in which this aspect may negatively impact direct or indirect users, and propose a way to measure this impact. In particular, we believe that the latter requirement may help steer authors toward considering more plausible impacts.

\paragraph{Reproducibility}
The reproducibility section of the card combines elements of \citet{mitchell2019modelcards}'s model cards and \citet{dodge-etal-2019-show}'s reproducibility checklist. The sections ask the minimal number of questions which are key to reproducing the model submission.
We request a model description, which includes the model type, version, the environment (i.e., versions of required software), and training algorithm used, with available space for further details. We also request a specification of dependencies and external libraries used to build the model. Authors have the option to link to their source code. Finally, authors are asked to describe the compute infrastructure used (e.g., the number of GPUs, the GPU type, and vRAM) and the training time for the final model.

Several questions concern the model hyperparameters, including the optimizer, training steps, learning rate. In addition, we elicit information about potential hyperparameter searches conducted as part of the model development. The hyperparameter search section requests information on the bounds for the hyperparameters, the number of search trials, and the method for choosing the hyperparameter values. The hyperparameter specifications for the best performing models are also requested but not required. 
Finally, the section ends with an optional space to list the number of training and evaluation runs and a required subsection detailing the utilized training dataset(s), including any processing on the data.

\paragraph{Evaluation}
The final section consists of the evaluation description. We suggest summarizing the evaluation process by including the metric details, the splits for the training, validation, and test data, and by providing the model performance on the test and validation data.

\section{Conclusion} \label{sec:conclusion}

We detail the processes and principles we followed to produce the documentation templates used in the HF library and the GEM benchmark. We ground this work in the current discussion of documentation as a way to communicate the impacts of ML systems. 
As touched on in Section~\ref{sec:methods-approach}, extensive documentation is only a tool to support the communication of decisions that led to the creation of datasets and models and the positionality of their creators; it is not a direct solution to the harms caused by ML systems.
We present our templates to encourage others to consider important questions that may be asked of their own work. The templates from both case studies are open source and we welcome contributions and feedback from authors and users to continually revise and improve them. Moreover, while the templates described in this work are designed for specific contexts and may not be fully applicable to others, they can be used as starting points for adaption to other settings.

\section{Social Impact and Positionality Statement} \label{sec:impact-and-positionality}

\paragraph{Social impact statement}
Our goal is to promote the standardization of specialized documentation for NLP datasets and models. Institutional adoption and promotion may see its greatest effect in the widening of community engagement. The infrastructure used to host and maintain the documentation also facilitates revisions and smaller contributions from the involved communities.
However, we are also aware of the risks that requiring this level of documentation for participation in either of our organizations may produce, such as raising the barrier to entry for those without experience in writing such documentation for language data as well as for people with fewer mentoring resources or platforms for engaging the community. 
Finally, while documenting the limitations of a resource is an important first step towards incrementally addressing issues, there is a risk that the act of documenting may allow creators to abdicate responsibility for these limitations in some cases, without taking any further steps to minimize negative social impacts of the systems they develop.

\paragraph{Positionality statement}
We are researchers at academic and industrial institutions with backgrounds in linguistics, NLP, ML, and HCI. Our guiding principles are discussed in Section \ref{sec:methods}. We aim to adapt available schemata to our specialized contexts, namely the HF library and the GEM benchmark and to present our development process as part of the general progress towards accountable and practical documentation for language datasets in ML systems. As NLP practitioners, we developed these card templates to directly support other members of the HF and NLG communities in writing documentation that answers questions that we ourselves would ask about the data and models. The completed templates will support users, researchers, and members of the public who may be impacted by these resources in understanding their contents and context.

\section*{Acknowledgments}

Thank you to our anonymous reviewers and colleagues for their thoughtful comments on this work.

\bibliographystyle{acl_natbib}
\bibliography{anthology,acl2021}

\appendix

\onecolumn

\section{Example of a Hugging Face Data Card: ELI5} \label{sec:appendix-eli5}

\subsection{Dataset Description}

\href{https://facebookresearch.github.io/ELI5/explore.html}{ELI5 homepage}; \href{https://github.com/facebookresearch/ELI5}{ELI5 repository}; paper: \href{https://arxiv.org/abs/1907.09190}{ELI5: Long Form Question Answering}; contact: Yacine Jernite

\paragraph{Dataset Summary}
The ELI5 dataset is an English-language dataset of questions and answers gathered from three subreddits where users ask factual questions requiring paragraph-length or longer answers. The dataset was created to support the task of open-domain long form abstractive question answering (QA), and covers questions about general topics in its \href{https://www.reddit.com/r/explainlikeimfive/}{r/explainlikeimfive} subset, science in it \href{https://www.reddit.com/r/askscience/}{r/askscience} subset, and history in its \href{https://www.reddit.com/r/AskHistorians/}{r/AskHistorians} subset.

\paragraph{Supported Tasks and Leaderboards}
The dataset can be used to train a model for Open Domain Long Form QA. An LFQA model is presented with a non-factoid and asked to retrieve relevant information from a knowledge source (such as \href{https://www.wikipedia.org/}{Wikipedia}), then use it to generate a multi-sentence answer. The model performance is measured by how high its \href{https://huggingface.co/metrics/rouge}{ROUGE} score to the reference is. A \href{https://huggingface.co/yjernite/bart_eli5}{BART-based model} with a \href{https://huggingface.co/yjernite/retribert-base-uncased}{dense retriever} trained to draw information from \href{https://huggingface.co/datasets/wiki_snippets}{Wikipedia passages} achieves a \href{https://yjernite.github.io/lfqa.html#generation}{ROUGE-L of 0.149}.

\paragraph{Languages}
The dataset is in English (BCP-47 code: \verb|en|), as spoken by users of the target subreddits.

\subsection{Dataset Structure}

\paragraph{Data Instances}
A typical data point comprises a question, with a \verb|title| containing the main question and a \verb|selftext| which sometimes elaborates on it, and a list of answers from the forum sorted by the number of upvotes they obtained. The URLs in each of the text fields have been  extracted to respective lists and replaced by generic tokens in the text. 
Examples are available \href{https://huggingface.co/datasets/viewer/?dataset=eli5}{here}.


\paragraph{Data Fields}
\verb|q_id|: a unique string question ID, corresponding to its ID in the source submission dumps; \verb|subreddit|: the source subreddit- `explainlikeimfive', `askscience', or `AskHistorians'; \verb|title|: title of the question, with URLs extracted and replaced by tokens in the form \verb|URL_n|; \verb|title_urls|: list of the extracted URLs, the nth element of the list was replaced by \verb|URL_n|; \verb|selftext|: either an empty string or an elaboration of the question; \verb|selftext_urls|: similar to \verb|title_urls|, but for \verb|self_text|; \verb|answers|: a list of answers, each answer has: \verb|a_id| (a unique string answer ID, corresponding to its ID in the source comments dumps), \verb|text| (the answer text with the URLs normalized), and \verb|score| (the number of upvotes the answer had received when the dumps were created); \verb|answers_urls|: a list of the extracted URLs (All answers use the same list, the numbering of the token continues across answer texts).

\paragraph{Data Splits}

The data is split into a training, validation and test set for each of the three subreddits. In order to avoid having duplicate questions in across sets, the \verb|title| field of each of the questions were ranked by their tf-idf match to their nearest neighbor and the ones with the smallest value were used in the test and validation sets. The number of training, validation, and test examples for each subreddit are: 272,634, 9,812, and 24,512 for \href{https://www.reddit.com/r/explainlikeimfive/}{r/explainlikeimfive}; 131,778, 2,281, and 4,462 for
\href{https://www.reddit.com/r/askscience/}{r/askscience}; and 98,525, 4,901, and 9,764 for
\href{https://www.reddit.com/r/AskHistorians/}{r/AskHistorians}. 

\subsection{Dataset Creation}

\paragraph{Curation Rationale}

ELI5 was built to provide a testbed for machines to learn how to answer more complex questions, which requires them to find and combine information in a coherent manner. The dataset consists of questions that were asked by community members of three subreddits, including \href{https://www.reddit.com/r/explainlikeimfive/}{r/explainlikeimfive}, and the answers provided by other users. The \href{https://www.reddit.com/r/explainlikeimfive/wiki/detailed_rules}{rules of the subreddit} make this data well-suited for abstractive QA: the questions need to seek an objective explanation about well established facts, and the answers provided need to be understandable without any particular domain knowledge.

\paragraph{Source Data: Initial Data Collection and Normalization}

The data was obtained by filtering submissions and comments from the subreddits of interest from the XML dumps of the \href{https://www.reddit.com/}{Reddit forum} hosted on \href{https://files.pushshift.io/reddit/}{Pushshift.io}.
In order to further improve the quality of the selected examples, only questions with a score of at least 2 and at least one answer with a score of at least 2 were selected for the dataset. The dataset questions and answers span a period form August 2012 to August 2019.

\paragraph{Source Data: Who are the source language producers?}

The language producers are users of the \href{https://www.reddit.com/r/explainlikeimfive/}{r/explainlikeimfive}, \href{https://www.reddit.com/r/askscience/}{r/askscience}, and \href{https://www.reddit.com/r/AskHistorians/}{r/AskHistorians} subreddits between 2012 and 2019. No further demographic information was available from the data source.

\paragraph{Annotations}

The dataset does not contain any additional annotations.



\paragraph{Personal and Sensitive Information}

The authors removed the speaker IDs from the \href{https://files.pushshift.io/reddit/}{Pushshift.io} dumps but did not otherwise anonymize the data. Some of the questions and answers are about contemporary public figures or individuals who appeared in the news.

\subsection{Considerations for Using the Data}

\paragraph{Social Impact of Dataset}

The purpose of this dataset is to help develop better question answering systems.
A system that succeeds at the supported task would be able to provide a coherent answer to even complex questions requiring a multi-step explanation, which is beyond the ability of even the larger existing models. The task is also thought as a test-bed for retrieval model which can show the users which source text was used in generating the answer and allow them to confirm the information provided to them.
It should be noted however that the provided answers were written by Reddit users, an information which may be lost if models trained on it are deployed in down-stream applications and presented to users without context. The specific biases this may introduce are discussed in the next section.

\paragraph{Discussion of Biases}

While Reddit hosts a number of thriving communities with high quality discussions, it is also widely known to have corners where sexism, hate, and harassment are significant issues. See for example the \href{https://www.reddit.com/r/announcements/comments/gxas21/upcoming_changes_to_our_content_policy_our_board/}{recent post from Reddit founder u/spez} outlining some of the ways he thinks the website's historical policies have been responsible for this problem, \href{https://www.researchgate.net/publication/283848479_Gamergate_and_The_Fappening_How_Reddit's_algorithm_governance_and_culture_support_toxic_technocultures}{Adrienne Massanari's 2015 article on GamerGate} and follow-up works, or a \href{https://www.wired.com/story/misogyny-reddit-research/}{2019 Wired article on misogyny on Reddit}.
While there has been some recent work in the NLP community on \textit{de-biasing} models (e.g. \href{https://arxiv.org/abs/1904.04047}{Black is to Criminal as Caucasian is to Police: Detecting and Removing Multiclass Bias in Word Embeddings} for word embeddings trained specifically on Reddit data), this problem is far from solved, and the likelihood that a trained model might learn the biases present in the data remains a significant concern.
We still note some encouraging signs for all of these communities:  \href{https://www.reddit.com/r/explainlikeimfive/}{r/explainlikeimfive}  and  \href{https://www.reddit.com/r/askscience/}{r/askscience}  have similar structures and purposes, and  \href{https://www.reddit.com/r/askscience/}{r/askscience}  was found in 2015 to show medium supportiveness and very low toxicity when compared to other subreddits (see a \href{https://hackerfall.com/story/study-and-interactive-visualization-of-toxicity-in}{hackerfall post},  \href{https://www.thecut.com/2015/03/interactive-chart-of-reddits-toxicity.html}{thecut.com write-up} and supporting \href{https://chart-studio.plotly.com/~bsbell21/210/toxicity-vs-supportiveness-by-subreddit/#data}{data}). Meanwhile, the \href{https://www.reddit.com/r/AskHistorians/wiki/rules}{r/AskHistorians rules} mention that the admins will not tolerate ``\textit{racism, sexism, or any other forms of bigotry}". However, further analysis of whether and to what extent these rules reduce toxicity is still needed.
We also note that given the audience of the Reddit website which is more broadly used in the US and Europe, the answers will likely present a Western perspectives, which is particularly important to note when dealing with historical topics.

\paragraph{Other Known Limitations}

The answers provided in the dataset represent the opinions of Reddit users. While these communities strive to be helpful, they should not be considered to represent a ground truth.

\subsection{Additional Information}

\paragraph{Dataset Curators}

The dataset was initially created by Angela Fan, Ethan Perez, Yacine Jernite, Jason Weston, Michael Auli, and David Grangier, during work done at Facebook AI Research (FAIR).

\paragraph{Licensing Information}

The license hinges on the legal status of the \href{https://files.pushshift.io/reddit/}{Pushshift.io} data which is unclear.

\paragraph{Citation Information} The citation can be found in the \href{https://www.aclweb.org/anthology/P19-1346/}{ACL Anthology}.

\paragraph{Contributions}

Thanks to \href{https://github.com/lewtun}{@lewtun}, \href{https://github.com/lhoestq}{@lhoestq}, \href{https://github.com/mariamabarham}{@mariamabarham}, \href{https://github.com/thomwolf}{@thomwolf}, and \href{https://github.com/yjernite}{@yjernite}.
\section{Example of a GEM Data Card: ASSET} \label{sec:appendix-asset}

\subsection{Dataset Description}

\href{https://github.com/facebookresearch/asset}{ASSET repository}; paper: \href{https://www.aclweb.org/anthology/2020.acl-main.424.pdf}{ASSET: A Dataset for Tuning and Evaluation of Sentence Simplification Models with Multiple Rewriting Transformations}; contact: Fernando Alva-Manchego, Louis Martin

\paragraph{Dataset and Task Summary}

\href{https://github.com/facebookresearch/asset}{ASSET} (\href{https://www.aclweb.org/anthology/2020.acl-main.424.pdf}{Alva-Manchego et al., 2020}) is multi-reference dataset for the evaluation of sentence simplification in English. The dataset uses the same 2,359 sentences from \href{https://github.com/cocoxu/simplification/}{TurkCorpus} (\href{https://www.aclweb.org/anthology/Q16-1029.pdf}{Xu et al., 2016}) and each sentence is associated with 10 crowdsourced simplifications. Unlike previous simplification datasets, which contain a single transformation (e.g., lexical paraphrasing in TurkCorpus or sentence splitting in \href{https://www.aclweb.org/anthology/D18-1081.pdf}{HSplit}), the simplifications in ASSET encompass a variety of rewriting transformations.

\paragraph{Why is this dataset part of GEM?}

ASSET is a high quality simplification dataset where each source (not simple) sentence is associated with 10 human-written simplifications. It is one of the two datasets for the text simplification task in GEM. It acts as the validation and test set.

\paragraph{Languages}

ASSET contains English text only (BCP-47: \verb|en|).

\subsection{Meta Information}

\paragraph{Dataset Curators}

ASSET was developed by researchers at the University of Sheffield, Inria, Facebook AI Research, and Imperial College London. The work was partly supported by Benoît Sagot's chair in the PRAIRIE institute, funded by the French National Research Agency (ANR) as part of the ``Investissements d’avenir" program (reference ANR-19-P3IA-0001).

\paragraph{Licensing Information}

\href{https://creativecommons.org/licenses/by-nc/4.0/}{Attribution-NonCommercial 4.0 International (CC BY-NC 4.0)}

\paragraph{Citation Information} The citation can be found in the \href{https://www.aclweb.org/anthology/2020.acl-main.424}{ACL Anthology}.


\paragraph{Leaderboard}

There is no official leaderboard associated with ASSET.

\subsection{Dataset Structure}

\paragraph{Data Instances}

\verb|simplification| configuration: an instance consists of an original sentence and 10 possible reference simplifications; \verb|ratings| configuration: an instance consists in an original sentence, an automatically generated simplification, and a human judgment of quality along one of three axes.
    
\paragraph{Data Fields}

\verb|original|: an original sentence from the source datasets; \verb|simplifications|: in the simplification config, a set of crowdsourced reference simplifications; \verb|simplification|: in the ratings config, an automatically generated simplification of the original; \verb|aspect|: in the ratings config, how the simplification is evaluated (meaning, fluency, or simplicity); \verb|rating|: a quality rating between 0 and 100

\paragraph{Data Statistics}

ASSET does not contain a training set; many models use \href{https://github.com/XingxingZhang/dress}{WikiLarge} (Zhang and Lapata, 2017) for training. For GEM, \href{https://github.com/chaojiang06/wiki-auto}{Wiki-Auto} will be used for training the model.
Each input sentence has 10 associated reference simplified sentences. The statistics of ASSET are given below.
For the input sentences, the validation set has 2000 instances and the test set has 359, for a total of 2359 sentences. Therefore, for the validation set there are 20000 simplifications and for the test set there are 3590 simplifications for a total of 23,590 simplified sentences.
The test and validation sets are the same as those of \href{https://github.com/cocoxu/simplification/}{TurkCorpus}. The split was random.
There are 19.04 tokens per reference on average (lower than 21.29 and 25.49 for TurkCorpus and HSplit, respectively). Most (17,245) of the referece sentences do not involve sentence splitting.

\subsection{Dataset Creation}

\paragraph{Curation Rationale}

ASSET was created in order to improve the evaluation of sentence simplification. It uses the same input sentences as the \href{https://github.com/cocoxu/simplification/}{TurkCorpus} dataset from (\href{https://www.aclweb.org/anthology/Q16-1029.pdf}{Xu et al., 2016}). The 2,359 input sentences of TurkCorpus are a sample of ``standard" (not simple) sentences from the \href{https://www.informatik.tu-darmstadt.de/ukp/research_6/data/sentence_simplification/simple_complex_sentence_pairs/index.en.jsp}{Parallel Wikipedia Simplification (PWKP)} dataset (\href{https://www.aclweb.org/anthology/C10-1152.pdf}{Zhu et al., 2010}), which come from the August 22, 2009 version of Wikipedia. The sentences of TurkCorpus were chosen to be of similar length (\href{https://www.aclweb.org/anthology/Q16-1029.pdf}{Xu et al., 2016}). No further information is provided on the sampling strategy.
The TurkCorpus dataset was developed in order to overcome some of the problems with sentence pairs from Standard and Simple Wikipedia: a large fraction of sentences were misaligned, or not actually simpler (\href{https://www.aclweb.org/anthology/Q16-1029.pdf}{Xu et al., 2016}). However, TurkCorpus mainly focused on \textit{lexical paraphrasing}, and so cannot be used to evaluate simplifications involving \textit{compression} (deletion) or \textit{sentence splitting}. HSplit (\href{https://www.aclweb.org/anthology/D18-1081.pdf}{Sulem et al., 2018}), on the other hand, can only be used to evaluate sentence splitting. The reference sentences in ASSET include a wider variety of sentence rewriting strategies, combining splitting, compression and paraphrasing. Annotators were given examples of each kind of transformation individually, as well as all three transformations used at once, but were allowed to decide which transformations to use for any given sentence.
An example illustrating the differences between TurkCorpus, HSplit and ASSET is given below:

\noindent \textit{Original}: He settled in London, devoting himself chiefly to practical teaching.

\noindent \textit{TurkCorpus}: He rooted in London, devoting himself mainly to practical teaching.

\noindent \textit{HSplit}: He settled in London. He devoted himself chiefly to practical teaching.

\noindent \textit{ASSET}: He lived in London. He was a teacher.

\paragraph{Communicative Goal}

The goal is to communicate the main ideas of source sentence in a way that is easier to understand by non-native speakers of English. This could be done by replacing complex words with simpler synonyms (i.e. paraphrasing), deleting unimportant information (i.e. compression), and/or splitting a long complex sentence into several simpler ones.

\paragraph{Source Data: Initial Data Collection and Normalization} 
Data from \href{https://github.com/cocoxu/simplification/}{TurkCorpus} (\href{https://www.aclweb.org/anthology/Q16-1029.pdf}{Xu et al., 2016})

\paragraph{Source Data: Who are the source language producers?}
The dataset uses language from English Wikipedia (August 22, 2009 version): some demographic information is provided \href{https://en.wikipedia.org/wiki/Wikipedia:Who_writes_Wikipedia\%3F}{here}.

\paragraph{Annotations: Annotation process}
The instructions given to the annotators are available \href{https://github.com/facebookresearch/asset/blob/master/crowdsourcing/AMT_AnnotationInstructions.pdf}{here}.

\paragraph{Annotations: Who are the annotators?}

Reference sentences were written by 42 workers on Amazon Mechanical Turk (AMT). The requirements for being an annotator were: (1) passing a qualification test (appropriately simplifying sentences), (2) being a resident of the US, UK or Canada, (3) having a HIT approval rate over 95\%, and over 1000 HITs approved.
Out of 100 workers, 42 passed the qualification test.
No other demographic or compensation information is provided in the ASSET paper.

\paragraph{Personal and Sensitive Information}

Since the dataset is created from English Wikipedia (August 22, 2009 version), all the information contained in the dataset is already in the public domain.

\subsection{Changes to the Original Dataset for GEM \normalfont No change.}

\subsection{Considerations for Using the Data} 

\paragraph{Social Impact of the Dataset}

The dataset helps move forward the research towards text simplification by creating a higher quality validation and test dataset. Progress in text simplification in turn has the potential to increase the accessibility of written documents to wider audiences.

\paragraph{Impact on Underserved Communities}

The dataset is in English, a language with many resources.

\paragraph{Discussion of Biases}

The dataset may contain some social biases, as the input sentences are based on Wikipedia. Studies have shown that the English Wikipedia contains both gender biases (\href{https://research.tudelft.nl/en/publications/is-wikipedia-succeeding-in-reducing-gender-bias-assessing-changes}{Schmahl et al., 2020}) and racial biases (\href{https://journals.sagepub.com/doi/pdf/10.1177/2378023118823946}{Adams et al., 2019}).

\paragraph{Other Known Limitations}

The dataset is limited to a small subset of topics present on Wikipedia.

\subsection{Getting started with in-depth research on the task}

The dataset can be downloaded from the original repository (\href{https://github.com/facebookresearch/asset}{here}) or be used via \href{https://huggingface.co/datasets/asset}{HuggingFace} and \href{https://www.tensorflow.org/datasets/overview}{TFDS}.
Recent supervised (\href{https://arxiv.org/abs/1910.02677}{Martin et al., 2019}, \href{https://www.aclweb.org/anthology/N19-1317/}{Kriz et al., 2019}, \href{https://www.aclweb.org/anthology/P19-1331/}{Dong et al., 2019}, \href{https://www.aclweb.org/anthology/D17-1062/}{Zhang and Lapata, 2017}) and unsupervised (\href{https://arxiv.org/abs/2005.00352v1}{Martin et al., 2020}, \href{https://www.aclweb.org/anthology/2020.acl-main.707/}{Kumar et al., 2020}, \href{https://www.aclweb.org/anthology/P19-1198/}{Surya et al., 2019}) text simplification models can be used as baselines.
A common metric for automatic evaluation is SARI (\href{https://www.aclweb.org/anthology/Q16-1029/}{Xu et al., 2016}).

\end{document}